\documentclass[fleqn]{annalen}
\usepackage{graphics}
\begin{document}
\newcommand{\volume}{8}              
\newcommand{\xyear}{1999}            
\newcommand{\issue}{5}               
\newcommand{\recdate}{29 July 1999}  
\newcommand{\revdate}{dd.mm.yyyy}    
\newcommand{\revnum}{0}              
\newcommand{\accdate}{dd.mm.yyyy}    
\newcommand{\coeditor}{ue}           
\newcommand{\firstpage}{507}         
\newcommand{\lastpage}{510}          
\setcounter{page}{\firstpage}        
\newcommand{\keywords}{disordered Fermi liquids, quantum magnetism, 
                                                   superconductivity}
\newcommand{\PACS}{71.10.-w; 75.30.Ds; 74.20.-z;}
\newcommand{\shorttitle}{D. Belitz et al., Disordered Fermi liquids}
\title{The disordered Fermi-liquid fixed point and its instabilities}
\author{D. Belitz$^{1}$ and T. R. Kirkpatrick$^{2}$} 
\newcommand{\address}
{$^1$Dept. of Physics and Materials Science Institute,
                                      University of Oregon, Eugene, OR 97403\\
$^2$Institute for Physical Science and Technology, and Department
             of Physics, University of Maryland, College Park, MD 20742\\}
\newcommand{\email}{\tt belitz@greatwhite.uoregon.edu} 
\maketitle
\def\tr{{\rm tr}\,}
\def\Tr{{\rm Tr}\,}
\def\sgn{{\rm sgn\,}}
\def\b{\bibitem}
\def\boldphi{\mbox{\boldmath $\phi$}}
\def\boldvarphi{\mbox{\boldmath $\varphi$}}
\begin{abstract}
The concept of a disordered Fermi-liquid fixed point is introduced and used
to understand various properties of disordered metals within a unifying
framework. Corrections to
scaling near this fixed point give what are commonly called 
weak-localization effects. Various instabilities of the disordered 
Fermi-liquid
phase are discussed. These include two distinct types of
superconducting-to-normal-metal quantum phase transitions.
First, the quantum phase
transition from a disordered metal to a conventional superconductor in bulk
materials is considered. Second, a quantum phase transition in
two dimensions from metallic-like behavior to a novel type of disorder
induced, spin-triplet, even-parity, superconductivity is treated. The paper is
concluded with a discussion of the nature of the ground state of
two-dimensional, disordered electron systems. 
\end{abstract}

\section{ Introduction}
\label{sec:I}

The many-fermion problem has a long history due to its importance in
condensed matter physics. Historically, this problem has been studied using
either many-body perturbation theory \cite{NegeleOrland}, or Landau's 
Fermi-liquid theory \cite{BaymPethick}. In
recent years various aspects of the many-fermion problem have been examined
using renormalization group (RG) and field theoretic techniques. Much
of this work has been done on clean fermion systems, applying the RG approach
either directly to the Grassmann field theory for fermions \cite{Shankar}, 
or to a
composite field theory that describes boson-like excitations in a fermion
system \cite{MarstonHoughton}. 
Among the results of this new approach are a RG derivation of
Fermi-liquid theory starting from a microscopic theory, and the introduction
of the Fermi-liquid fixed point notion. Other results include RG derivations
of the Cooper instability problem and of the random-phase approximations
that lead to, e.g., screening. 

For disordered electronic systems the RG approaches developed for clean
Fermi systems are not directly applicable, because the sets of soft modes
are very different for the two cases. However, a field-theoretic
description based on composite variables that are bilinear in the basic
fermionic fields, and hence effectively bosonic in nature, has been
used for some time. This approach has been pioneered by Wegner \cite{Wegner}, 
who showed
that, for noninteracting electrons, the effective field theory takes the form 
of a nonlinear sigma model. This was generalized by 
Finkel'stein \cite{F} to the case 
of interacting electrons. The key idea underlying these
effective theories is to keep explicitly only those degrees of freedom that
are likely to be relevant for the problem under consideration, and to
integrate out all others in some simple approximation.

In this paper we show how field theoretic and
RG methods can be used to describe the disordered
fermion problem. In particular we first review how Finkel'stein's theory can
be derived from a microscopic model with the aid of RG ideas. To this end we
introduce the notion of a disordered Fermi-liquid fixed point (FP), and show 
that corrections to scaling near this FP give what are commonly called 
weak-localization effects. We then discuss various instabilities of the
disordered Fermi-liquid fixed point. Physically, these instabilities indicate
phase transitions from the disordered Fermi-liquid phase to various magnetic
and superconducting metallic phases, as well as to magnetic and nonmagnetic
insulator phases.

\section{Fermionic field theory}
\label{sec:II}

\subsection{Grassmannian field theory}
\label{subsec:II.A}

Since the description of fermions involves anticommuting variables, any
field-theory description for electrons must be formulated in terms of
anticommuting or Grassmann variables \cite{NegeleOrland}. 
For simplicity we consider here a
model for a homogeneous electron fluid subject to a random potential that
models the quenched disorder. The partition function can then be written, 
\begin{equation}
Z=\int D[\bar \psi ,\psi ]\ \exp [S]\quad.  
\label{eq:2.1}
\end{equation}
Here the functional integration is with respect to Grassmann valued fields, 
$\bar \psi $ and $\psi $, and the action $S$ is given by 
\begin{equation}
S=S_0 + S_{\rm dis} + S_{\rm int}\quad.  
\label{eq:2.2}
\end{equation}
$S_0$ describes free electrons 
\begin{equation}
S_0 = \int dx\ \sum_\sigma \bar \psi _\sigma (x)\left[ -\partial _\tau 
      +\frac{\nabla ^2}{2m}+\mu \right]\,\psi_\sigma(x)\quad,
\label{eq:2.3}
\end{equation}
with $x=(\bf{x},\tau )$, and $\int dx=\int d{\bf x}\int_0^{1/T}d\tau$. 
$\bf{x}$ denotes position, $\tau $ imaginary time, $\sigma $ is a spin
label, $\mu $ is the chemical potential, and $m$ is the electron mass. 
$S_{\rm dis}$ describes a static random potential, $u(\bf{x})$, that couples
to the fermionic number density, 
\begin{equation}
S_{\rm dis}=-\int dx\ u(\bf{x})\sum_\sigma \bar \psi_\sigma(x)\,\psi _\sigma
(x)\quad,  
\label{eq:2.4}
\end{equation}
and $S_{\rm int}$ denotes a spin-independent two-particle interaction.

For disordered systems, in contrast to clean ones,
single-particle momentum eigenstates are not long lived, but
decay exponentially on a time scale given by the elastic mean-free time
for electron-impurity interactions. The important physics on the
longest length and times scales is controlled by two-particle excitations
that are soft either because they are related to conserved quantities or
because of a mechanism related to Goldstone's theorem. For a detailed
discussion of this point we refer the reader to Ref. \cite{us_fermions}. 
It turns out that of all the excitations whose interactions are described by 
$S_{\rm int}$, the dominant
soft modes are those that involve fluctuations of either the particle number
density $n_n$, or the spin density ${\bf n}_s$, or density fluctuations
in the Cooper channel, $n_c$. In Fourier space, $q=({\bf q},\omega_n)$,
with $\omega_n$ a Matsubara frequency, these densities can be easily
written in terms of fermion variables. The interaction, $S_{\rm int}$, can be
written in terms of these fluctuations as, 
\begin{equation}
S_{\rm int}=S_{\rm int}^{(s)}+S_{\rm int}^{(t)}+S_{\rm int}^{c(s)}\quad,  
\label{eq:2.5}
\end{equation}
with singlet, triplet, and Cooper channel terms denoting interactions
between the densities discussed above, 
\begin{equation}
S_{\rm int}^{(s)}=-\frac{\Gamma ^{(s)}}2\sum_q n_n(q)\,n_n(-q)\quad,  
\quad S_{\rm int}^{(t)}=\frac{\Gamma^{(t)}}{2}\sum_q{\bf n}_s(q)\cdot{\bf n}_s%
(q)\quad,  
\label{eq:2.6a}
\end{equation}
\begin{equation}
S_{\rm int}^{c(s)}=-\frac{\Gamma ^{(c)}}2\sum_q n_c(q)\,n_c(q)\quad.  
\label{eq:2.6b}
\end{equation}
Here $\Gamma ^{(i)}$, $i=s,t,c(s)$, are interaction amplitudes in the
particle-hole spin-singlet and triplet, and in the particle-particle
or Cooper spin-singlet interactions channels, respectively. Notice that
there is no Cooper spin-triplet interaction term. The reason is that for
models with static, point-like, interaction amplitudes, 
as we have assumed here, such
a term is forbidden by the Pauli principle. We will come back to this
in Sec.\ \ref{subsubsec:III.B.2} below.

\subsection{ Nonlinear sigma model}
\label{subsec:II.B}

The idea behind the nonlinear sigma model description of disordered electron
systems is that the long-distance (long-wavelength) and long-time 
(low-frequency) properties of the system are 
determined by the massless modes or
excitations. The derivation of an effective theory then becomes a two-step
process. First, the massless modes in the system need to be identified. 
For example, in a simple Fermi liquid with no long-range magnetic
order, the only soft modes are diffusive modes representing, e.g.,
number density, energy density, and spin density diffusion. Second, the
microscopic field theory, Eqs.\ (\ref{eq:2.2}) - (\ref{eq:2.6b}), 
must be transformed into an effective one that keeps only these
soft modes, while all other degrees of freedom are integrated out in some
approximation that respects crucial features of the full theory like, e.g.,
conservation laws. All of this can be done, 
and has been discussed in
great detail in Ref. \cite{us_fermions}. The resulting effective field theory 
is called a generalized
nonlinear sigma model. Within this theory the
partition function can be written in terms of matrix fields $Q$, 
\begin{equation}
Z=\int D[Q]\ e^{{\cal A}[Q]}\quad.  
\label{eq:2.7}
\end{equation}
The matrix elements of $Q$ are isomorphic to bilinear products of the
elements of $\bar\psi$ and $\psi$.
The action ${\cal A}$ is given by, 
\begin{equation}
{\cal A} = -\frac 1{2G}\int d{\bf x}\,\tr({\bf\nabla}Q({\bf x}))^2
    +2H\int d{\bf x}\ \tr(\Omega Q({\bf x})) + {\cal A}_{\rm int} \quad,
\label{eq:2.8}
\end{equation}
with ${\cal A}_{\rm int}$ given by Eqs.\ (\ref{eq:2.6a},\ref{eq:2.6b}) 
written in terms of $Q-$matrices. The
explicit derivation \cite{us_fermions}
yields the coupling constants as $G=8/\pi\sigma_0$,
with $\sigma_0$ the conductivity in the self-consistent Born approximation,
and $H=\pi N_F/8$ which can be interpreted as the 
quasiparticle density of states ($N_F$ is the bare single-particle density 
of states at the Fermi level) \cite{us_R}.
$G$ is a measure of the disorder in the system. $\Omega $ is a diagonal
matrix whose elements are the fermionic Matsubara frequencies $\omega_n$. 
$Q$ is subject to the constraints 
\begin{equation}
Q^2({\bf x})=1\quad,\quad Q^{\dagger }=Q\quad,\quad \tr\,Q({\bf x})=0\quad. 
\label{eq:2.9}
\end{equation}
A standard way to enforce these constraints is to write 
\begin{equation}
Q=\left( 
\begin{array}{cc}
(1-qq^{\dagger })^{1/2} & q \\ 
q^{\dagger } & -(1-q^{\dagger }q)^{1/2}
\end{array}
\right)  
\label{eq:2.10}
\end{equation}
where the matrix $q$ has elements $q_{nm}\equiv q(\omega_n,\omega_m)$ 
with $n\geq 0$, $m<0$.\footnote{We are suppressing some technicalities here.
For instance, the $q$ also carry replica indices, and the matrix elements are
quaternion-valued.}

The effective action given above is the generalized nonlinear sigma model
that was first proposed by Finkel'stein \cite{F} as a model for disordered
interacting electrons near a metal-insulator transition.

\subsection{ The disordered Fermi-liquid fixed point}
\label{subsec:II.C}

Insight into a disordered metal can be gained from studying the RG and fixed
point properties of the generalized sigma model field theory given by
Eq.\ (\ref{eq:2.8}). We begin our discussion of the model
by performing a momentum-shell RG procedure. For the rescaling part
of this transformation, we need to assign a scale dimension to the 
field $q$. Choosing the scale dimension of a length $L$ to be $[L]=-1$, 
we write 
\begin{equation}
[q({\bf x})]=\frac{1}{2} (d-2+\eta)\quad,  
\label{eq:2.11}
\end{equation}
which defines the exponent $\eta $. The stable Fermi-liquid FP of the theory
is characterized by the choice\footnote{We adopt the RG philosophy of S.-K.
Ma \cite{Ma}, where FPs are selected by choosing the values of certain
scale dimensions or critical exponents.}
\begin{equation}
\eta =0\quad.  
\label{eq:2.12}
\end{equation}
Physically this corresponds to diffusive correlations of $q$ in the
disordered Fermi-liquid phase. In addition, we must specify the scale
dimension of the frequency or temperature, i.e., the dynamical scaling
exponent $z=[\omega ]=[T]$. In order for the FP to be consistent with
diffusion, that is with frequencies that scale as the square of the
wavenumber, we must choose 
\begin{equation}
z=2\quad.  
\label{eq:2.13}
\end{equation}
Now we expand the sigma model action in powers of $q$. In a symbolic
notation that leaves out everything not needed for power counting purposes,
we write 
\begin{equation}
{\cal A}=\frac{-1}G\int d{\bf x}\,(\nabla q)^2 + H\int d{\bf x}\,\omega\,q^2 
   + \Gamma\,T\int d{\bf x}\,q^2 + O(\nabla^2q^4,\omega q^4,Tq^3)\quad,  
\label{eq:2.14}
\end{equation}
with $G\sim 1/\sigma_0$ and $H\sim N_F$, and 
$\Gamma $ denotes any of the interaction amplitudes. Power counting shows
that with the above choices for the exponents, all these coupling constants
have vanishing scale dimensions with respect to our FP: 
\begin{equation}
[G] = [H] = [\Gamma] = 0\quad.  
\label{eq:2.15}
\end{equation}
These terms (together with some others \cite{us_fermions} that we can
suppress for our present purposes) therefore make up our FP action.

Now consider the leading corrections to the FP action, as indicated in 
Eq.\ (\ref{eq:2.14}). Power counting shows that
all these terms are irrelevant to the Fermi-liquid FP as long as 
$d>2$, and they become marginal in $d=2$ and relevant for $d<2$. All
terms that were neglected in deriving the sigma model can be shown to be
even more irrelevant than the ones considered here.
We conclude that Eq.\ (\ref{eq:2.14}) is a FP action and that any of 
the leading irrelevant operators (which we collectively denote by $u$)
have dimension $[u]=-(d-2)$. We can use these results to discuss the
leading corrections to scaling near the disordered Fermi-liquid FP for
physical quantities such as the conductivity, the specific heat, and
various susceptibilities.

Let us first consider the dynamical conductivity $\sigma (\omega )$. Its
bare value is proportional to $1/G$, and according to Eq.\ (\ref{eq:2.15}) 
its scale dimension is zero. We therefore have the scaling law 
\begin{equation}
\sigma (\omega ) = \sigma (\omega b^z,ub^{-(d-2)})\quad,  
\label{eq:2.18}
\end{equation}
where b is an arbitrary RG scale factor. By putting $b=1/\omega ^{1/z}$, and
using $z=2$, Eq.\ (\ref{eq:2.13}), as well as the fact that 
$\sigma (1,x)$ is an
analytic function of x, we find that the conductivity has a singularity at
zero frequency, or a long time tail, of the form 
\begin{equation}
\sigma (\omega) = {\rm const} + \omega^{(d-2)/2}\quad.  
\label{eq:2.19}
\end{equation}
This nonanalyticity is well known from perturbation theory for both
noninteracting and interacting electrons \cite{LeeRama}. 
Note that these results were derived
basically without doing any calculations, illustrating the power
of the RG. The RG derivation also proves that the 
$\omega^{(d-2)/2}$ is the {\it exact} leading nonanalytic behavior.

Similarly, the low-frequency (or low-temperature, or small-wavenumber)
behavior of other physical quantities can be obtained. For example, the
single-particle density of states behaves as 
\begin{equation}
N(\omega\rightarrow 0) = {\rm const} + \omega^{(d-2)/2}\quad,  
\label{eq:2.20}
\end{equation}
and, as a function of wavenumber, the static spin susceptibility scales as 
\begin{equation}
\chi_s({\bf q},T=0) = {\rm const} + \left\vert{\bf q}\right\vert^{(d-2)}\quad.  
\label{eq:2.21}
\end{equation}
Here we have omitted the prefactors of the nonanalyticities.

\subsection{ Metal-insulator transitions}
\label{subsec:II.D}
In addition to the stable Fermi-liquid FP discussed in the previous 
subsection, there also is a critical FP that describes a metal-insulator
transition. This is the FP first discussed by Finkel'stein \cite{F}, and
later analyzed in greater detail by others. For lack of space we cannot
discuss this here, and refer the reader to the reviews \cite{us_R} 
and \cite{us_March}.

\section{Instabilities of the Fermi liquid}
\label{sec:III}

The Fermi-liquid FP discussed in Sec.\ \ref{sec:II}, and the phase it
describes, are perturbatively stable for weak interactions. However,
if the electron-electron interaction in some channel becomes large,
the Fermi-liquid FP becomes unstable, and the system
undergoes a transition to a different phase. We will be concerned with
magnetic and superconducting phases. The properties of quantum phase
transitions to magnetic phases are discussed elsewhere in these 
proceedings \cite{TV_talk,hh_islands}. However, before we turn to 
superconducting
instabilities, we will discuss some soft-mode induced properties of
itinerant ferromagnets that are akin to the weak-localization effects
discussed in Sec.\ \ref{sec:II} above.

\subsection{Magnons at zero temperature}
\label{subsec:III.A}

An important manifestation of quantum long-range order in a ferromagnet is
the existence of spin waves. In conventional Heisenberg ferromagnets the
damping of the spin waves is negligible, and in the long-wavelength limit
the spin-wave dispersion is
\begin{equation}
\Omega =D(m)\,{\bf q}^2\quad.  
\label{eq:3.1}
\end{equation}
The coefficient $D(m)$ depends on the dimensionless magnetization $m$. In
the conventional theory for clean `weak' ferromagnets \cite{Moriya}, 
$D(m\rightarrow 0) = D_0\,m$, with $D_0$ a constant. 
In general this result is not correct for
itinerant ferromagnets, because the same effects that lead to the 
weak-localization nonanalyticities discussed above also lead to a nonanalytic 
dependence of $D(m)$ on $m$ at $m=0$. 

To see this we use a scaling argument. For more details, as well as explicit
calculations, we refer the reader to Ref. \cite{us_magnons}. 
First we note that the coefficient $D$ in
Eq.\ (\ref{eq:3.1}) can be related to $m$ times the coefficient of 
the ${\bf q}^2$ term in the wavenumber expansion of a spin susceptibility. 
In the paramagnetic state, Eq.\ (\ref{eq:2.21}) 
shows that the leading small wavenumber term is actually 
${\bf q}^2\,\left\vert{\bf q}\right\vert^{d-4} = 
\left\vert{\bf q}\right\vert^{d-2}$. (Here we consider disordered systems,
we will come back to the clean case later.)
In the ferromagnetic state, on the other hand, the small
wavenumber singularity is cut off by the magnetization $m$. In particular, a
finite magnetization acts like a magnetic field in this respect and supplies
a frequency scale that is proportional to $m$ (viz., the cyclotron
frequency). For diffusive dynamics, the
wavenumber scales like the square root of the frequency. This implies a
length scale cutoff, $\ell_m$, due to the magnetization, that scales as 
$\ell_m\sim m^{-1/2}$. The net result is that for disordered systems, for
$2<d<4$, $D(m)$ behaves as
\begin{equation}
D(m\rightarrow 0) \propto m\,\ell_m^{4-d} \propto m^{(d-2)/2}\quad,
\label{eq:3.2}
\end{equation}
This leads to the striking experimental
prediction that in bulk systems, $D(m\rightarrow 0)\propto m^{1/2}$ compared
to the linear $m$-dependence predicted by conventional theories. The net
result is that the `speed' of spin waves in `weak' ($m\rightarrow 0$)
ferromagnets is larger than the conventional theory predicts.

For clean itinerant ferromagnets a similar, albeit weaker, effect has been
predicted \cite{us_magnons}. For bulk systems, 
$D(m\rightarrow 0)\propto m\,\ln (1/m)$, while
in two-dimensions, $D(m\rightarrow 0)\propto m^0 = {\rm const}$.

\subsection{Superconducting phases and phase transitions}
\label{subsec:III.B}

A large interaction in the Cooper channel triggers an instability of the
Fermi liquid in favor of a superconducting state. We first discuss the phase
transition to a conventional (BCS)
superconducting state at zero temperature. Then we discuss
how disorder can lead to more exotic types
superconductivity, especially in $d=2$.

\subsubsection{Superconducting-to-normal-metal transition at zero temperature}
\label{subsubsec:III.B.1}

The standard starting point to describe any phase transition is to construct
a Landau-Ginzburg-Wilson (LGW) functional for the order parameter
fluctuations. The basic idea is that these
fluctuations are the important ones to describe the large scale changes
associated with the phase transition. In a disordered metal the situation is
more complicated because, in addition to the order parameter fluctuations
being soft, the diffusive modes discussed in Sec.\ \ref{sec:II} 
are also soft, and in general
they couple to the order parameter fluctuations.
The usual procedure of integrating out all modes other than the order
parameter fluctuations therefore in particular integrates out some soft
modes, which leads to a nonlocal LGW theory.
Nevertheless, the critical behavior of
the resulting theory can still be determined
using RG ideas. Furthermore, since the nonlocality of the LGW functional
corresponds to a long-ranged effective interaction between the order
parameter fluctuations, the critical behavior can be determined {\it exactly}
from a Gaussian FP, as has been explained in more detail in
Ref.\ \cite{us_sc}. 

To illustrate this, we denote
the LGW functional by $\Phi $ and the superconducting order parameter by 
$\Psi $. Gauge invariance implies that only even powers of $\Psi $ appear 
in $\Phi $. The coefficients in this expansion, i.e., the vertex functions of
the effective field theory, are connected correlation functions of the
anomalous or Cooper channel density $n_c$, see Eq.\ (\ref{eq:2.6b}), 
in a system away from
any phase transition point \cite{Hertz}. 
This auxiliary system we will refer to as
a reference ensemble. In particular, the
Gaussian vertex is determined by the anomalous density-density correlation
in the reference ensemble. Denoting the latter by $C(q)$, the Gaussian term
in the LGW functional is, 
\begin{equation}
\Phi ^{\left( 2\right) }[\Psi ]=\int dq\,\Psi ^{*}(q)\left[ 1/\left| \Gamma
_c\right| -C(q)\right] \Psi (q)\quad.
\label{eq:4.1}
\end{equation}
$C(q)$ is a complicated correlation function. However, since the reference 
ensemble is a disordered Fermi liquid, its structure is known \cite{us_R}.
RG arguments show that the structure of $C$ at low
frequencies and long wavelengths in the limit $T\rightarrow 0$ is 
\begin{equation}
C(q)=\frac Zh\frac{\ln (\Omega_0/(D{\bf q}^2+\left| \Omega _n\right| ))}{%
1+(\delta \Gamma ^{c(s)}/h)\ln (\Omega_0/(D{\bf q}^2+\left| \Omega
_n\right| ))}\quad.  
\label{eq:4.2}
\end{equation}
Here $\Omega_0$ is a frequency cutoff on the order of the Debye frequency
(for phonon-mediated superconductivity), and $\delta \Gamma ^{c(s)}$ is a
repulsive interaction in the Cooper channel that is generated in
perturbation theory even if its bare value is zero. $D$ is the diffusion
coefficient of the electrons in the reference system, $h$ is the
renormalized value of $H$ (see Eq.\ (\ref{eq:2.8})), 
and $Z$ is a wavefunction renormalization. Using
these results in Eq.\ (\ref{eq:4.1}) 
we obtain a Gaussian LGW functional that has the form
\begin{equation}
\Phi ^{(2)}=\int dq\,\Psi ^{*}(q)\left[ t+\frac 1{\ln (\Omega_0/(D{\bf q}%
^2+\left| \Omega _n\right| ))}\right] \Psi (q)\quad,  
\label{eq:4.3}
\end{equation}
where $t=-(Z\left\vert\Gamma^{c(s)}\right\vert - \delta\Gamma^{c(s)})$ is the
bare distance from the phase transition.

From the structure of the Gaussian vertex $\Gamma^{(2)}$, i.e., the term in
brackets in Eq.\ (\ref{eq:4.3}),
we read off the values of the
exponents $\eta$, $\gamma$, and $z$, defined as $\Gamma ^{(2)}({\bf q}%
,\Omega =0)\sim \left| {\bf q}\right| ^{2-\eta },\Gamma ^{(2)}({\bf q}%
=0,\Omega =0)\sim t^\gamma ,$ and $\xi _\tau \sim \xi ^z$, with $\xi _\tau $
the relaxation time. They are 
\begin{equation}
\eta =2\quad,\quad\gamma =1\quad,\quad z=2\quad.  
\label{eq:4.4}
\end{equation}
By scaling $\left\vert \bf{q}\right\vert $ 
with the correlation length $\xi $,
we also obtain the scaling behavior of the latter, 
\begin{equation}
\xi \sim \exp (1/2\left\vert t\right\vert )\quad.  
\label{eq:4.5}
\end{equation}
The exponent $\nu $ as usually defined therefore does not exist, 
$\nu =\infty $.

In order to determine how the order parameter behaves near the critical
point, one needs the quartic term $\Phi ^{(4)}$ in the LGW functional. 
One finds that the exponent $\beta $ also
does not exist, $\beta =\infty $, but that the ratio $\beta /\nu $ is finite
and is equal to the expected (from scaling) value 
\begin{equation}
\beta /\nu =2\quad.  
\label{eq:4.6}
\end{equation}
Finally, it is also possible to determine the behavior of the conductivity
and specific heat near the quantum transition point. For these details we
refer to Ref. \cite{us_sc}.

\subsubsection{Novel superconducting states: disordered-induced spin-triplet
even-parity superconductivity}
\label{subsubsec:III.B.2}

As mentioned after Eq.\ (\ref{eq:2.6b}), the Pauli principle does not
allow for a static, point-like interaction amplitude,
$\Gamma^{c(t)}$, in the particle-particle spin-triplet channel.
More generally, any point-like $\Gamma^{c(t)}$ must be an odd function
of frequency. In general, this means that
if $\Gamma^{c(t)}$ exists,
then it vanishes at zero frequency, and is 
therefore not important at long times
or low temperatures. This seems to preclude spin-triplet even-parity
superconductivity triggered by $\Gamma^{c(t)}$. However, this argument is
fallacious for low-dimensional disordered electron systems. The basic point 
is that in the presence of electronic interactions the diffusive modes that
lead to weak-localization effects, and to nonlocal theories for magnetic and
other superconducting phase transitions, also lead to a $\Gamma^{c(t)}$ 
with a nonanalytic frequency dependence that approaches a discontinuity
as $d\rightarrow 2$. As a result, the presence of both
disorder and electron-electron interactions seem to inevitably {\it lead} to
exotic superconductivity in $d=2$!

To understand physically how this happens let us first consider the effect
of the spin-triplet interaction amplitude in the {\it particle-hole}
channel, $\Gamma^{(t)}$. By means of $\Gamma^{(t)}$,
an electron spin polarizes its environment.
This polarization is ferromagnetic. Suppose an electron has created a spin
polarization cloud and then moves away. In a clean system the polarization
cloud at a given point will decay quickly in time. The polarization cloud
therefore essentially moves with the electron that creates it, giving rise
to the standard Fermi-liquid renormalization of the spin susceptibility. 
In a disordered
system, a spin density fluctuation $\delta n_s$ will decay algebraically
like $\delta n_s\sim t^{-d/2}$, because of the long-time tail effects
already discussed (see Eq.\ (\ref{eq:2.19})). 
The polarization cloud will therefore persist even after
the electron that created it has diffused away. A second electron moving
into the region at a later time will still see the remains of the
polarization cloud. It will get attracted to it if the two electrons form a
spin triplet. We therefore expect an attractive contribution of $\Gamma
^{(t)}$ to $\Gamma^{c(t)}$ .

Next consider the effect of the spin-singlet interaction, $\Gamma ^{(s)}$.
By means of $\Gamma ^{(s)}$, an electron charge polarizes its environment.
Again, while in a clean system this leads only to Fermi-liquid
renormalizations, in a disordered system the resulting charge polarization
cloud will decay as $\delta n_c\sim t^{-d/2}$. A second electron will be
attracted to this region regardless of spin. The conclusion is that both
$\Gamma^{(t)}$ and $\Gamma^{(s)}$ will lead to an attractive $\Gamma^{c(t)}$.
Further, since the Fourier transform of $t^{-d/2}$ is proportional to
$\ln\omega$ in $d=2$, we expect these contributions to be singular at low
frequencies, or low temperature. Finally, since there is no bare repulsive
$\Gamma ^{c(t)}$
to overcome, the conclusion is that in $d=2$ there is a tendency
toward a novel type of superconductivity at $T=0$.

Explicit calculations confirm the above physical picture \cite{us_R}. 
Further, both the
critical properties near this quantum phase transition and the
physical properties of the novel superconducting state have been studied
in some detail. For the phase transition properties we refer 
to Ref. \cite{us_tsc}. Here
we restrict ourselves to briefly discussing some interesting properties
of the superconducting state.

The first relevant question concerns the (mean-field) critical temperature
for reasonable parameter values. As noted above, there are
two mechanisms leading to superconductivity. For realistic MOSFET parameters,
the charge fluctuation mechanism leads to $T_c \approx 10\,{\rm mK}$.
The spin fluctuation mechanism, on
the other hand, can lead to a mean-field $T_c$ that is a fraction of the
Fermi temperature, i.e., about $1K$. Keeping in mind that mean-field
estimates generally overestimate $T_c$, we conclude that 
in at least some
systems, this type of superconductivity is a possibility.
Other quantities of interest are the tunneling density of states, the
diamagnetic susceptibility, and the conductivity. Model
calculations yield a density of states that vanishes logarithmically
as the bias voltage approaches zero, so the predicted superconducting
state has a pseudo-gap. Calculations have also established that 
there is a conventional Meissner
effect, and that the electrical conductivity
is identical in form to that in a conventional BCS
superconductor. These results answer some questions that had been raised
about the stability of odd-gap superconductivity. For details, see
Ref. \cite{us_tsc}.

\section{The two-dimensional ground state problem}
\label{sec:IV}

Our discussion so far implies that the nature of the ground state of a
generic $2$-$d$, interacting, disordered electron system is not known,
and that there are a number of possibilities. Indeed, this is one of the
most fundamental open problems in quantum many-body physics, 
and we call it the two-dimensional ground state problem.

For noninteracting electrons, at zero magnetic
field, the ground state is known to be an Anderson insulator
\cite{LeeRama}. When repulsive
electron-electron interactions are added, a number of distinct possibilities
arise. First, it could still be an insulator. Indeed, this was the
conventional view until very recently. Second, the ground state
could be the novel disorder-induced superconducting state discussed above.
Third, it has been known for some time that if there are no magnetic
impurities or impurity spin orbit scattering, there is a tendency toward 
some type of magnetic state that may
(or may not) be conducting \cite{us_R}. Fourth, there have been speculations
that some other, 
even more exotic, state may occur \cite{non-FL}.

Experimentally, the situation is very interesting. There is mounting
evidence that in some $2$-$d$ systems, especially low-electron density
MOSFETs, there is some kind of a
conducting phase at very low temperatures \cite{Kravchenko}.
Other $2$-$d$ systems are surely insulators. 
These observations are consistent with the
notion that there are numerous possibilities for the $2$-$d$ ground state,
and that the ultimate choice
depends on microscopic details and cannot be universally determined. A
detailed free energy calculation will be needed to determine what ground
state is the thermodynamically stable one for a specific system. 

\vspace*{0.25cm} \baselineskip=10pt{\small\noindent
We would like to thank our collaborators on some of the magnetic topics
discussed above, Andy Millis, Rajesh Narayanan, and Thomas Vojta, and the
Aspen Center for Physics for hospitality during the completion of this
paper. This work
was supported by the NSF under grant numbers DMR-98-70597 and DMR-99-75259.}

\end{document}